\documentclass[superscriptaddress,twocolumn,amsmath,amssymb,aps,pre,english]{revtex4-2}

\usepackage{dcolumn}
\usepackage{bm}
\usepackage{ulem} 
\usepackage{inputenc}
\usepackage[T1]{fontenc}
\usepackage{upgreek,babel,calc,mathrsfs}
\usepackage{amsmath,amssymb,graphicx,xcolor}
\usepackage[export]{adjustbox}
\usepackage[unicode=true]{hyperref}
\begin{document}

\title{Sorting of multiple molecular species on cell membranes}

\author{Andrea Piras}
\thanks{These Authors contributed equally.}
\affiliation{Candiolo Cancer Institute, FPO - IRCCS, 
{str.~prov.~142}, km 3.95, 
10060 Candiolo, Italy}
\affiliation{Institute of Condensed Matter Physics and Complex Systems,
Department of Applied Science and Technology, Politecnico di Torino,
Corso Duca degli Abruzzi 24, 
10129 Torino, Italy}
\affiliation{Italian Institute for Genomic Medicine (IIGM),
{str.~prov.~142}, km 3.95, 
10060 Candiolo, Italy}
\affiliation{Department of Oncology, University of Turin, 10060 Candiolo, Italy}
\author{Elisa Floris}
\thanks{These Authors contributed equally.}
\affiliation{Institute of Condensed Matter Physics and Complex Systems,
Department of Applied Science and Technology, Politecnico di Torino,
Corso Duca degli Abruzzi 24, 
10129 Torino, Italy}
\affiliation{Istituto Nazionale di Fisica Nucleare (INFN), 
Sezione di Torino, Via Pietro Giuria 1, 10125 Torino,
Italy}
\author{Luca Dall'Asta}
\email{luca.dallasta@polito.it}
\affiliation{Institute of Condensed Matter Physics and Complex Systems,
Department of Applied Science and Technology,
Politecnico di Torino, Corso Duca degli Abruzzi 24, 10129 Torino, Italy}
\affiliation{Italian Institute for Genomic Medicine (IIGM), 
{str.~prov.~142}, km 3.95, 
10060 Candiolo, Italy}
\affiliation{Istituto Nazionale di Fisica Nucleare (INFN), 
Sezione di Torino, Via Pietro Giuria 1, 10125 Torino,
Italy}
\affiliation{Collegio Carlo Alberto, Piazza Arbarello 8, 10122, Torino, Italy}
\author{Andrea Gamba}
\email{andrea.gamba@polito.it}
\affiliation{Institute of Condensed Matter Physics and Complex Systems,
Department of Applied Science and Technology, Politecnico di Torino,
Corso Duca degli Abruzzi 24, 10129 Torino, Italy}
\affiliation{Italian Institute for Genomic Medicine (IIGM), 
{str.~prov.~142}, km 3.95, 
10060 Candiolo, Italy}
\affiliation{Istituto Nazionale di Fisica Nucleare (INFN), 
Sezione di Torino, Via Pietro Giuria 1, 10125 Torino,
Italy}

\begin{abstract}
Eukaryotic cells maintain their inner order by a hectic process of distillation of molecular factors taking place on the
surface of their lipid membranes. To understand the properties of this molecular sorting process, 
a physical 
model 
of the process
has
been recently proposed~\cite{ZVS+21}, based on 
(a)~the~phase~separation of a single, initially dispersed 
molecular species into spatially localized sorting domains on the lipid membrane,~and 
(b)~domain-induced 
membrane bending 
leading to the
nucleation
of 
submicrometric
lipid vesicles,
naturally
enriched in the molecules of the engulfed sorting domain. 
The analysis of the model has 
shown the existence of 
an optimal region of the parameter space where sorting is most efficient. Here, 
the model is extended 
to account for the simultaneous distillation of a 
pool of 
distinct 
molecular species. 
We find that the mean time spent by sorted molecules  on the membrane 
 increases with the heterogeneity of the pool  
(i.e., 
the number of distinct molecular species sorted)
according to a simple scaling law, 
and that a large number of distinct molecular species can in principle 
be 
sorted in parallel on a typical cell membrane region without significantly interfering
with each other.
Moreover, sorting is found to be most efficient when the distinct molecular species have
comparable homotypic affinities.
We also consider how valence (i.e., the average number of  
interacting 
neighbors of a molecule
in a sorting domain) 
affects the sorting process, finding 
that 
higher-valence 
molecules 
can be 
sorted with greater efficiency than lower-valence molecules. 
\end{abstract}

\maketitle

\section{\label{sec:level1} Introduction}
Eukaryotic cells have developed 
a complex
mechanism of
molecular 
distillation,
allowing 
them 
to 
impart
\hbox{distinct} chemical identities 
to  
a variety of
inner
cell organelles 
that 
host 
specific 
sets 
of molecules and lipids
and 
perform
specific functions~\cite{MN08,SCC+12}.
At the 
basis 
of this process
of organelle maintenance and renewal
is a hectic 
traffic
of vesicles generated 
through a sophisticated
process of 
sorting of molecular factors 
taking place 
both
on the 
outer
cell membrane and on the membranes 
that enclose  
inner cell compartments. 
It 
has been recently proposed
that 
this  
fundamental biological
process
may emerge
from
the combination of two simple physical mechanisms~\cite{ZVS+21}:
(a)~the~phase~separation of a single, initially dispersed molecular species into 
spatially localized sorting domains,~and 
(b)~domain-induced 
nucleation
of 
submicrometric
lipid vesicles,
that become then
naturally
enriched in the molecules contained in the engulfed sorting domain. 
Based on these assumptions, a phenomenological theory was developed, where
the main control parameter is the intermolecular interaction
strength which drives the 
phase separation process 
and regulates the critical
size of  
sorting domains, 
separating 
small, transient, ``unproductive'' sorting 
domains from larger, ``productive'' 
sorting domains destined to grow and
to be ultimately extracted~\cite{ZVS+21,FPP+22}.
It is \hbox{important} to observe here that the interaction between homotypic molecules
leading to their phase separation into distinct sorting domains may be either
\textit{direct}, such as in the case of the weakly adhesive electrostatic interactions 
between unstructured molecule regions involved in biological liquid-liquid phase 
separation~\cite{BLH+17}, or \textit{indirect}, as in the case of the
effective, contactless
interactions 
induced by 
enzyme-driven 
feedback loops involving lipid and molecules.
The latter, indirect interactions 
have the potential to 
induce diffusion-limited phase
separation, originally studied in the context of cell 
polarity~\cite{GCT+05,GKL+07,SVN+12,HBF18,FPA+21}, 
and 
are
involved in 
molecular sorting processes where the segregation of distinct molecular species
in separate sorting domains is not controlled by direct homotypic intermolecular
interactions~\cite{CLS+20}.
In~our phenomenological approach, the \hbox{observable}
\textit{effective} \hbox{\textit{interaction}} \textit{strength} measures
the tendency of homotypic molecules to become enriched in localized spatial
regions, irrespective of the microscopic (direct or indirect) origin of the
attraction~\cite{ZVS+21,FPP+22}.

A natural 
measure of the efficiency of the 
distillation
process is 
the time a 
molecule
spends
on a given membrane region
before being sorted and extracted: 
the shorter this residence time, the larger the
distillation efficiency. 
In the steady state, 
the distillation process
is most efficient
for intermediate aggregation strengths, where both the 
molecular residence time 
and the average surface density of sorted
molecules are minimal, and  
both are
related to the 
incoming molecule flux 
through simple scaling
laws~\cite{ZVS+21}. 
The phenomenological theory 
reproduces well
the experimental 
distribution
of sorting domain sizes~\cite{ZVS+21}
and the experimental distributions of the lifetimes and 
maximum sizes of both productive and unproductive sorting domains~\cite{FPP+22, WCM+20}.   

In the past, experimental investigations have been
mainly focused on the process of sorting of 
single molecular species, such as transferrin receptors or low density
lipoproteins~\cite{MN08}. 
More recently, advances in \hbox{imaging} technologies have made it possible to 
elucidate 
aspects of the simultaneous
distillation of distinct molecular species, and to directly observe their
localization in distinct, separate sorting domains~\cite{RKS+20,RKS+21,STI+21,GM16}. 
The demixing of distinct molecular species subject to attractive homotypic interactions at equilibrium is predicted by statistical physics arguments~\cite{Des10,MD11}.
Here, we propose an extension of the 
nonequilibrium
theory of molecular sorting
introduced in Ref.~\onlinecite{ZVS+21}, where sorting of a single molecular species was
considered, 
to the case where a plurality of distinct molecular species is sorted in
parallel in the same membrane region.
In this more general case,
by combining theoretical arguments and 
lattice-gas
numerical simulations,
we show
that 
the mean time sorted molecules \hbox{reside} on the membrane 
increases with the heterogeneity of the pool
(i.e., the number of distinct molecular species sorted)
according to a simple scaling law, 
and that a large number of distinct molecular species can in principle 
be 
sorted in parallel on a typical cell membrane region without significantly interfering
with each other.
Moreover, sorting is found to be most efficient when 
distinct molecular species have
comparable homotypic affinities.
Since recent studies have highlighted a crucial role of valence
(i.e., the average number of interacting neighbors of a molecule in a crowded
homotypic domain) 
in driving phase separation 
and sorting
on cell membranes~\cite{LBC+12,BR14,BZJ+22}, 
we also performed numerical simulations of the sorting of 
molecules with different valence, finding that   
higher-valence 
molecules 
can be
sorted with greater efficiency. 

\section{\label{sec:theory} Phenomenological theory for the parallel sorting of multiple
molecular species}

The phenomenological theory
of 
molecular sorting 
driven by phase separation
previously introduced
in Ref.~\onlinecite{ZVS+21} is here
generalized to the case of 
the parallel distillation of
$N>1$ non-interacting molecular species.
The theory describes
``cargo'' molecules 
continuously injected into the lipid
membrane
in random positions,
and then 
laterally 
diffusing
on the membrane.
Attractive 
(direct or indirect)
interactions between 
homotypic
molecules 
(i.e., molecules belonging to 
the same species)
can
lead to the formation of
multiple 
homotypic
sorting domains enriched in 
the
molecules of 
a 
particular
species,
thus inducing 
a natural demixing process.
For~simplicity, 
the case where all the distilled species have similar
biophysical properties will be mainly considered.
In particular,  
each molecule will be assumed to occupy approximately the same 
characteristic area~$A_0$ in a sorting domain.  

Effective 
homotypic
attractive interactions drive the growth of 
approximately circular,
homotypic
sorting domains
by 
the
absorption of freely diffusing molecules, that form a sort of two-dimensional
``gas'' of molecules surrounding the growing domains. 
When a sorting domain reaches 
a characteristic 
area  
$A_E=m A_0$, 
it is
extracted from the membrane system
through the formation of a separate lipid vesicle, 
that thus becomes naturally enriched in the molecular species 
contained in the engulfed domain.
This self-organizing process of parallel distillation of $N$ molecular species
defines a non-equilibrium steady-state, whose statistical properties are
determined by the incoming flux of molecules and by the 
strength of the attractive  
interaction between homotypic molecules diffusing on the given membrane region. 
In the low density
regime, the process of formation of 
domains enriched 
in 
a specific type of
cargo molecule 
is approximately independent of 
the formation of domains 
of 
the 
other species, and is 
regulated by the 
value
of 
the  
critical size 
of sorting domains, beyond which irreversible
domain growth takes place~\cite{FPP+22}.

The growth of an
approximately circular domain of 
the $i$-th 
species 
is 
driven by 
the 
flux $\Phi_{i}$
of such molecules across 
the domain 
boundary. 
In the 
quasi-stationary regime,
and in the limit of 
low-density gas and
approximately absorbing
domains~\cite{ZVS+21},
$\Phi_{i}$
can be computed 
by 
solving
a two-dimensional Laplace equation with 
\hbox{Dirichlet} boundary
conditions
and circular symmetry, 
finding 
\begin{equation}
\Phi_i \sim D_i\,\bar{n}_i,\label{eq:PhiiisD}
\end{equation}
where $\bar{n}_i$ is the 
average 
density of freely diffusing molecules of 
the 
$i$-th
species, 
and $D_i$ 
is 
the corresponding 
diffusion
coefficient. 
In the stationary
regime, all particles of the $i$-th species injected in the system are eventually
absorbed by supercritical domains, therefore 
\begin{equation}
\phi_i \sim \Phi_i\, \bar{N}_{d,i}
\end{equation} 
where 
$\bar{N}_{d,i}$ is the average density of supercritical domains of the
$i$-th species.

In the non-equilibrium stationary state, the 
average number
$\phi_i$ of particles
of the $i$-th species 
injected into the membrane system per unit time and unit area
equals 
the 
analogous
number of such particles leaving the system as a consequence of domain
extractions. 
Under the assumption that supercritical domains
grow irreversibly until extraction, one gets 
\begin{equation} 
\phi_i = m\,\dfrac{d \bar{N}_{d, i}}{d t}, \label{eq:phiiism}
\end{equation} 
where ${d \bar{N}_{d, i}}/{d t}$
is the rate of formation of supercritical domains of the $i$-th species per unit
membrane area.
The~rate of formation of 
such
domains depends on the 
frequency of formation of germs of new sorting domains
and on the probability that those germs reach the supercritical stage,
and can 
be expressed 
phenomenologically~as  
\begin{equation}
    \dfrac{d \bar{N}_{d, i}}{d t} = C_i D_i \bar{n}^2_i, 
    \label{eq:domain_formation}
\end{equation}
where 
$C_i$ is a dimensionless 
quantity representing the macroscopic,
effective strength of the attractive interaction acting between homotypic
molecules~\cite{ZVS+21,FPP+22}. 
 
According to a general steady-state relation valid for open systems in a driven
non-equilibrium stationary state, the average density of particles in the system
is given by the product of the average density flux of particles and the average
residence time of a particle in the system~\cite{ZDG19}. 
In the present case, this relation can be applied to several entities that populate 
the membrane in the statistically stationary state.
For the total, average 
density 
$\rho$ of molecules 
(both freely diffusing and bound to sorting domains)
of all $N$ species
in the stationary state
one has 
\begin{equation}
\rho = \phi\, \bar{T}, 
\end{equation}
where $\bar{T}$ is 
the average molecule
residence time on the
membrane,
and $\phi=\sum_{i=1}^N \phi_i$. 
This shows in particular that for fixed values of the molecular flux $\phi$,
the average residence time $\bar T$ is simply proportional to the average 
molecule density $\rho$.
For
the average density 
of
freely diffusing molecules of the $i$-th species one~finds 
\begin{equation}
    \bar{n}_i =  \phi_i\, \bar{T}_{f,i},
\end{equation}
where $\bar{T}_{f,i}$ is the average time a 
molecule of the $i$-th species spends 
in the gas. 
For the average density of supercritical domains of the $i$-th species 
one has
\begin{equation}
    \bar{N}_{d,i} =  \dfrac{d \bar{N}_{d, i}}{d t}\, \bar{T}_{d,i}  =   \dfrac{\phi_i}{m}
     \,\bar{T}_{d,i},
    \label{eq:domaindensity}
\end{equation}
where $\bar{T}_{d,i}$ is the average 
lifetime of a sorting domain.
The latter, 
in the limit of approximately absorbing
domains, is of the order of the average
time a molecule of the $i$-th species
spends as a part of a sorting domain.  
For simplicity, we only analyze 
here
the symmetric
case, where 
$C_i = C$ and $D_i=D$ for all $i=1,\ldots,N$, and 
also assume that
$\bar{T}_{f,i}=\bar{T}_{f}$,  
\ $\bar{T}_{d,i}=\bar{T}_{d}$,
and $\phi_i=\phi/N$ for all $i=1,\ldots,N$.
The total density of molecules in the gas 
is then
\begin{equation}
\bar{n} =  \phi \,\bar{T}_{f},
\end{equation} 
while the total number of
supercritical domains 
per 
unit 
area 
is 
\begin{equation}
\bar{N}_{d} = \phi \,\bar{T}_{d}/m .\label{eq:ndisphi} 
\end{equation} 
Combining the 
relations (\ref{eq:PhiiisD}--\ref{eq:ndisphi}), all 
the main 
quantities describing the behavior
of the system in the non-equilibrium,
statistically stationary state can be 
\hbox{explicitly} 
expressed in terms
of the 
number of species $N$, 
the 
total 
incoming 
molecule
flux $\phi$, the extraction size $m=A_{E}/A_{0}$, the 
diffusivity~$D$, and the phenomenological 
interaction strength~$C$,~as: 
\begin{align} \bar{n} &
\sim \left( \dfrac{\phi\, N}{m\, C\, D}  \right)^{1 / 2} , \label{eq:gas_density} \\
\bar{N}_d & \sim \left( \dfrac{m\, C\, \phi\, N}{D}  \right)^{1 / 2}, \label{eq:domain_density} \\ 
\bar{T}_{d} & \sim \left (\dfrac{m^3\,C\, N }{D\, \phi}\right)^{1/2}, \label{eq:res_time} \\ 
\bar{T}_{f} & \sim \left( \frac{N}{m\, C\, D \,\phi }\right)^{1/2}. \label{eq:free_time}
\end{align}
The efficiency of the molecular sorting process 
in the steady state 
is inversely proportional to 
the 
mean
time of
residence 
of a cargo molecule
on the membrane, 
approximately given by
$\bar{T} =\bar{T}_{f} + \bar{T}_{d}$. 
The highest efficiency is obtained when 
$\bar{T}$
is 
minimal, i.e.,
for
\begin{equation}
C \sim  C_{\rm opt} \sim m^{-2}.
\end{equation}
In this optimal regime, 
each 
molecule spends approximately the same
amount of time freely diffusing in the gas and as a part of a growing domain,~i.e.
\begin{equation}
    \bar{T}_{d} \sim \bar{T}_{f} \sim  \left (\frac{m\, N}{D\, \phi} \right)^{1/2}.
    \label{eq:T_opt}
\end{equation} 
For 
fixed 
incoming flux $\phi$, 
the mean 
total 
residence time of a molecule on the
membrane increases 
therefore
with the number of different species as~$ \bar{T} \sim N^{1 / 2}$.
In the optimal regime,
the average density of freely diffusing molecules and 
the average density of 
supercritical
domains behave as  
\begin{align} \bar{n}_{\mathrm{opt}} & \sim \left (
\dfrac{m\, \phi \,N}{D} \right )^{1 / 2}, \\ 
\label{eq:Nd} \bar{N}_{d,\mathrm{opt}} & \sim \left (\dfrac{\phi\, N}{m\,D} \right )^{1/2}.
\end{align} 
From 
(\ref{eq:gas_density}) and (\ref{eq:domain_density}), 
it follows that also the total molecule density 
$\rho \sim \bar{n} + m\, \bar{N}_d$ 
scales as $\rho \sim N^{1/2}$, and 
is minimal
for
$C\sim C_{\rm opt}$.

A consequence of the above relations is 
that even for low, fixed values of the \textit{total} incoming 
molecule
flux $\phi$,
the low-density regime, 
where 
molecules and domains of different species
do not interact
significantly,  
progressively breaks down as the number $N$ of species increases. 
For very high $N$, the crowding of 
molecules of different species surrounding a domain 
can be expected to
cause 
an effective
decrease in the flux $\Phi_i$ at the surface, 
such that (\ref{eq:PhiiisD}) 
should
be modified into
$\Phi_i \propto  f_N D \,\bar{n}_i$, 
with 
$f_N$
a
decreasing 
function
of~$N$. 
In that 
case, a simple modification of the 
previous
phenomenological 
arguments 
gives 
$C_{\rm opt} \sim f_N \, m^{-2}$, 
thus predicting
that the optimal effective interaction between homotypic molecules
should 
decreases 
for very large $N$.

A rough indication 
about the number of
different species 
that
can be sorted in parallel 
without significantly interfering with each other 
in the optimal sorting regime 
can be obtained by 
noticing that in 
the 
low-density regime,
the interdomain half 
distance $L$ 
has to 
be much 
larger than the extraction 
size  
$A_E^{1/2}$. 
Observing that 
$\pi\, L^2 \bar{N}_d \sim 1$ and using Eq.~\ref{eq:Nd}, 
this condition translates into
\begin{equation}
  L \sim \bar{N}_d^{-1/2} \sim \left (\frac{D \,m}{N\, \phi} \right )^{1/4} \gg A_E^{1/2}
\end{equation}
giving
\begin{equation}
    N\ll\frac{D}{A_E A_{0}\phi} 
    \label{eq:estimate}
\end{equation}
Using the
realistic orders of magnitude 
$D\sim 10^{-3} \mathrm{\mu m}^2/\mathrm{s}$, 
$A_E\sim 10^{-1} \mathrm{\mu m}^2$, 
$A_0\phi\sim 10^{-5} \,\mathrm{s}^{-1}$ 
for the process of endocytic sorting~\citep{ZVS+21} 
one obtains 
$N\ll 10^3$.
Therefore, 
the simple analytical estimate (\ref{eq:estimate}),
based on the phenomenological theory of sorting, suggests 
that a large number of 
different molecular species may in principle be distilled in parallel on a
typical membrane system.
It is worth 
observing,
however,
that
Eq.~\ref{eq:estimate} 
can provide only a qualitative indication about the breakdown of
the low-density regime, as it was derived neglecting the contribution
of complicated logarithmic prefactors~\cite{ZVS+21}.
The onset of  
the
regime 
of molecular crowding will be
therefore
more precisely investigated in the next
Section by means of numerical simulations
of a lattice-gas implementation of the
sorting process.

The ordering effect of the sorting process on the molecule gas can be 
quantified as follows.
Consider the case where molecule injection takes place by 
the fusion 
into
a membrane system of area $A$
of
vesicles carrying a well-mixed cargo of molecules of all of the $N$ 
distinct
molecular species. 
In the steady state, the same 
number of cargo molecules
is extracted in 
vesicles 
containing in average
$\mathcal{N}_i=A\,\phi_i\,\delta t$ 
molecules of only one of the $i$-th molecular species.
Formally treating empty membrane regions of area $A_0$ as a 0-th molecular species, 
we define a corresponding
flux density $\phi_0$.
The change in entropy 
due to the demixing process can 
then
be 
measured~as~\cite{LL80,BSM+13}:
\begin{equation}
    \delta S = \sum_ {i=0}^N \mathcal{N}_i \log
    \frac{\mathcal{N}_i}{\mathcal{N}},
    \label{eq:mixing}
\end{equation}
(with $\mathcal{N}=\sum_{i=0}^N \mathcal{N}_i $)
resulting in a simple expression for
the 
average rate of entropy production per unit \hbox{membrane} area: 
\begin{eqnarray*}
    \frac{1}{A}\frac{\delta S}{\delta t} &=&
    \sum_{i=0}^N \phi_i \log \frac{\phi_i}{\phi_0+\phi} \\ &=& - \phi_0 \log \left( 1 + \frac{\phi}{\phi_0} \right) - \phi \log \left( 1 +
\frac{\phi_0}{\phi} \right)-\phi \log N. 
    \label{eq:entropy_prod}
\end{eqnarray*}

\section{Numerical Results}

\begin{figure}[tb]
\includegraphics[width=0.4\textwidth]{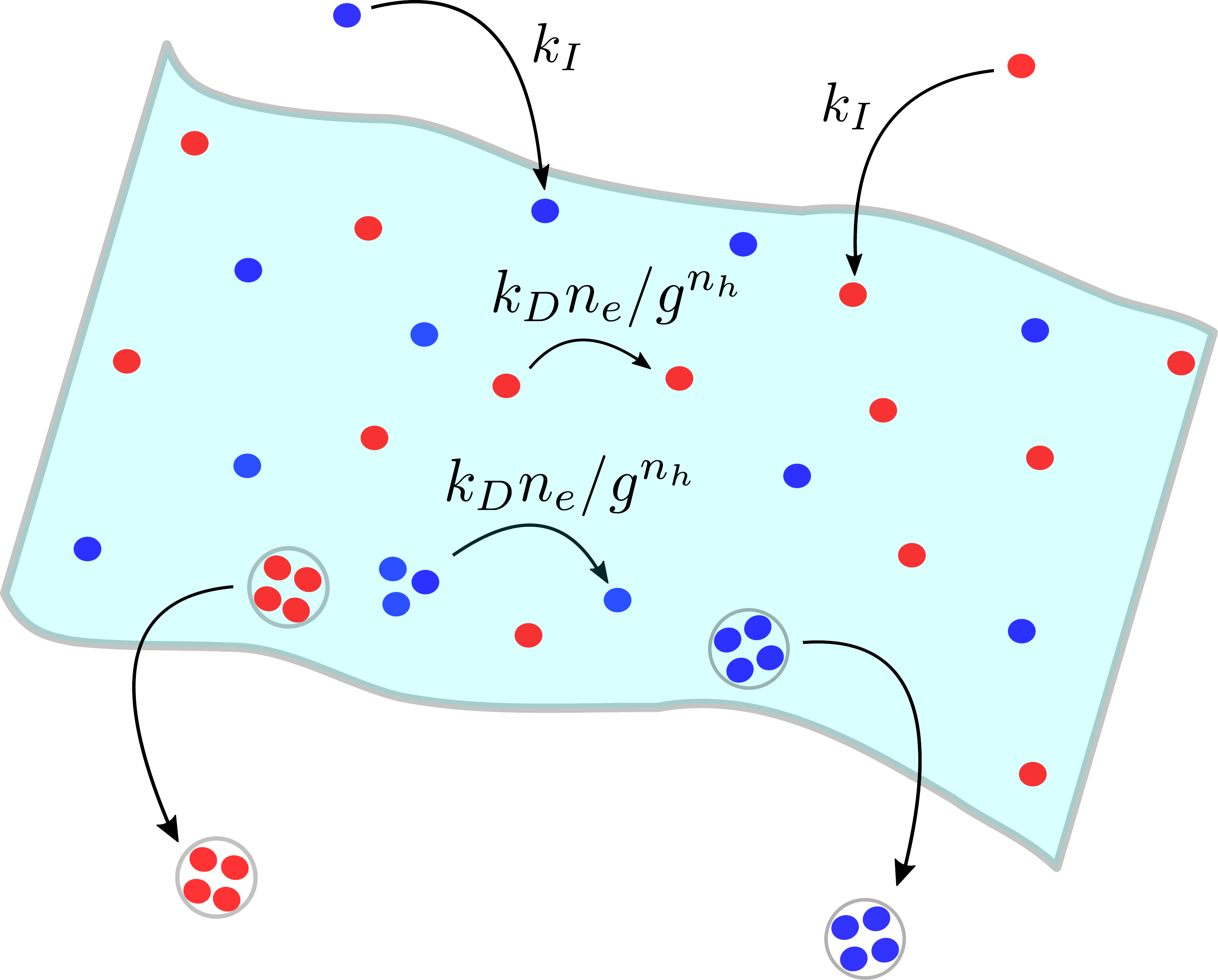}
  \caption{Schematic representation of the 
   hard-core
  lattice-gas model used 
  to investigate 
   the parallel
   sorting 
  of multiple molecular species.} 
  \label{fig:scheme}
  \end{figure}

The  
sorting
of
a plurality of
distinct molecular species is further investigated by means of numerical simulations of 
a generalized version of the hard-core lattice-gas model previously 
introduced
in Ref.~\onlinecite{ZVS+21}. 
A~schematic representation of the
stochastic processes 
taking place in
the model is shown in
Fig.~\ref{fig:scheme}.
Cargo
molecules 
are individually inserted on unoccupied
sites 
of a lattice
with rate~$k_I$ 
and 
diffuse  
by hopping on  
unoccupied
neighboring sites with rate $k_D$. 
The affinity between molecules of the same
species induces a decrease in the mobility rate $k_D$ by a factor $g^{n_h}$,
where $g$ is 
a microscopic measure of 
interaction strength, and $n_h$ is the number of
neighboring sites occupied by 
homotypic 
molecules. 
Molecules of
different species are 
inserted in empty lattice sites with the same rate,
and the only interaction between them is due to excluded volume effects.
Areas are measured as multiples of the area $A_0$ of a lattice site.
In~the stationary state, the incoming flux per site
is 
$\phi=k_I(1-\rho)$, where $\rho$ is the stationary 
molecule
density on the lattice. 
Connected homotypic domains that reach the size $m$ are extracted from the
system.
Further
details on 
the lattice-gas model are provided in
App.~\ref{sec:latticegas}. 

\begin{figure}[tb]
~\!\!\!\!\!\!\!\!\!\includegraphics[width=1.05\columnwidth]{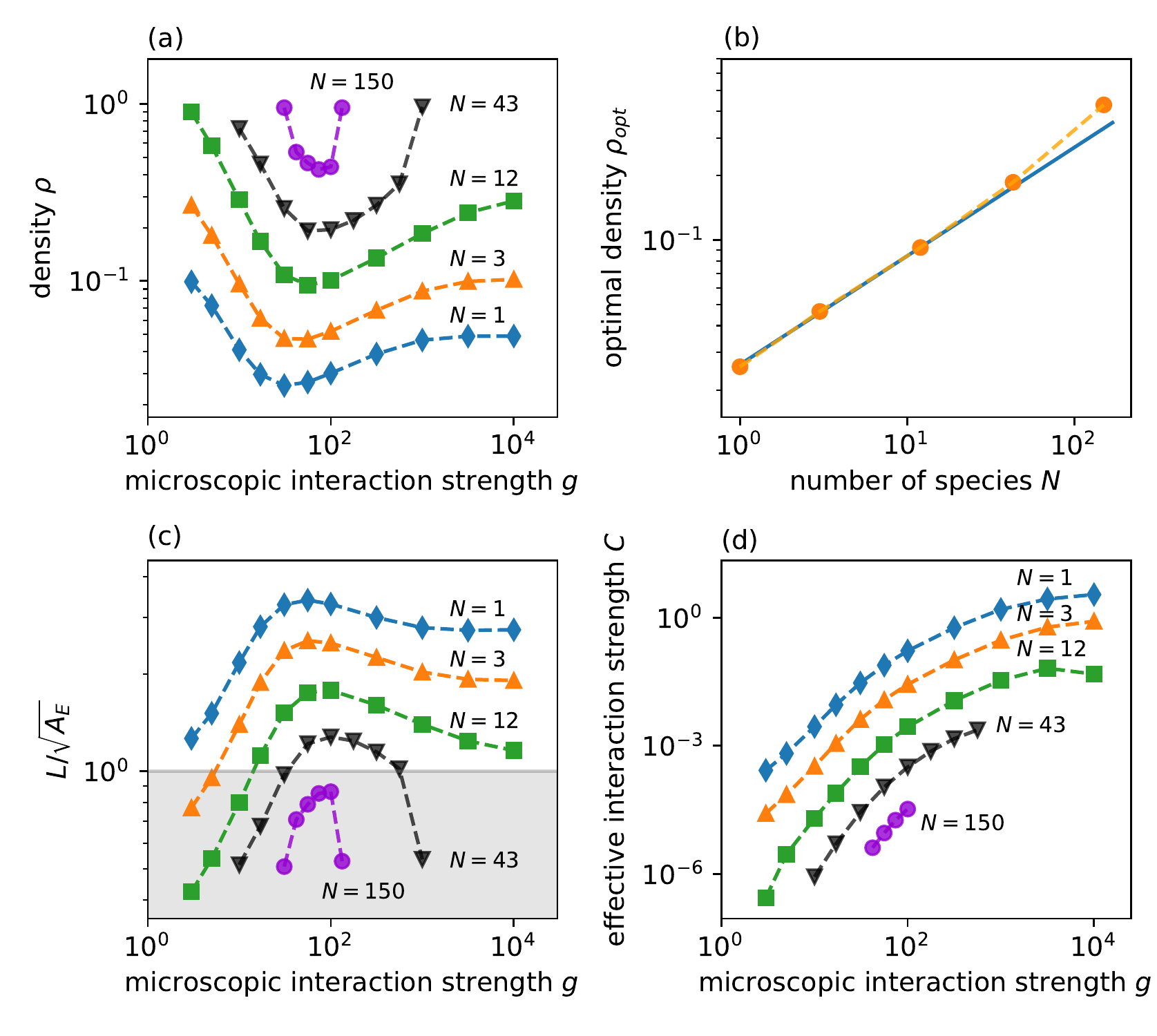}
  \caption{(a) Total molecule density as a function of the microscopic
  interaction strength $g$, for increasing values of the number of species~$N$.
 For large~$N$, sorting remains possible
 only
  in a restricted interval of values of $g$.
  (b)~Total 
  molecule
  density 
  as a function of the number of molecular
  species~$N$ in the regime of optimal sorting. The blue solid line is fitted
  with the law $\rho \sim N^a$, with $a =0.51$. This scaling relation 
  breaks down for $N\sim 10^2$.
  (c) Ratio of the interdomain half-distance~$L$ to the extraction 
size~$A_E^{1/2}$
as 
a function of the microscopic interaction strength $g$. 
The 
parameter
region 
where
the system 
becomes overcrowded
and sorting 
is impaired 
is marked in light gray.
(d) Effective interaction strength $C$ as a function of the microscopic
interaction strength $g$ for different values of the number of molecular species
$N$. Simulations were performed at fixed incoming molecule
  flux $\phi/k_D=10^{-5}$ on a square lattice.  } \label{fig:N} \end{figure}

\subsection{Sorting multiple molecular species}

The results of numerical simulations
of the model 
with  
the realistic flux 
$\phi/k_D =
10^{-5}$ 
are displayed in Fig.~\ref{fig:N}. 
Consistently with previous results~\cite{ZVS+21,FPP+22}, Fig.~\ref{fig:N}a shows
that, 
in 
an intermediate range of values of the microscopic interaction
strength $g$, the stationary density of molecules $\rho$ exhibits a minimum,
corresponding to an
optimal sorting regime. Increasing the number $N$ of 
sorted 
species, the
optimal region moves 
towards
larger values of $g$, indicating that efficient molecular
sorting becomes more 
and more 
difficult to obtain 
as the number of sorted species increases.
For 
very 
large $N$, molecular sorting 
can take place 
only in
a restricted interval of values
of $g$, 
as 
the system tends to freeze into an overcrowded state for 
both
lower  
and
higher 
values of $g$. 
The existence of a maximum value~$g_N$ such that sorting
at $g>g_N$ becomes impossible for
large~$N$  due to crowding effects can  
also 
be
checked 
by
looking at the behavior of
the stationary molecular density 
$\rho_{g=\infty}$ as a function of $N$ 
(App.~\ref{app:crowding}, Fig.~\ref{fig:g_inf}(a)): 
this
density 
rapidly transitions to values of order 1 
(high molecular crowding) for $N\sim 10$, 
signaling that sorting is strongly impaired at high $g$ for $N\gtrsim 10$.

\begin{figure}[b]
  \includegraphics[width=1\columnwidth]{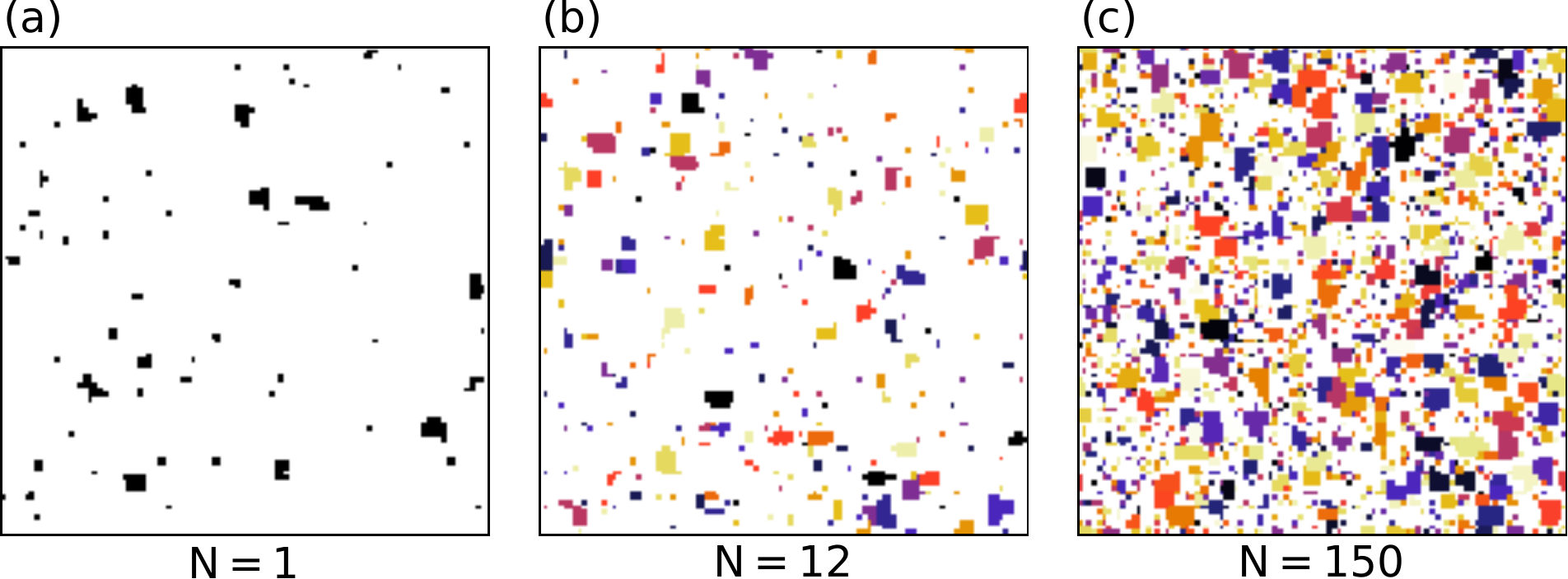}
  \caption{Snapshots of the sorting process for \hbox{increasing} \hbox{values} of the number of
  species $N$ (from left to right). 
  \hbox{Different} molecular species have been marked with different colors.
  Simulations were performed 
  in the optimal regime at the realistic value 
  $\phi/k_D=10^{-5}$
  of the 
  incoming molecule flux.     }
  \label{fig:snapN}
\end{figure}

When $N\sim 10$, 
sorting 
is still
possible 
around the optimal region, which however tends to shrink 
progressively
with increasing $N$,
since,
as seen in the previous \hbox{Section},
the optimal
molecular
density $\rho_\mathrm{opt}$ 
grows with $N$, in quantitative
agreement with the scaling law $\rho_\mathrm{opt} \sim N^{1/2}$ predicted by the
phenomenological theory (Fig.~\ref{fig:N}(b)). 
The deviation from the $N^{1/2}$ scaling observed at 
$N\sim10^2$
signals
the breakdown,
even at optimality, 
of the low-density regime where 
the 
parallel processes of 
sorting
of different species
take place
approximately independently
of each other.
For larger 
$N$,
sorting
domains are no longer well separated and 
molecular 
mobility 
is strongly
reduced (App.~\ref{app:crowding}, Fig.~\ref{fig:g_inf}(b)). 
The behavior of the interdomain half distance $L$, numerically evaluated as
a
function of $g$ for various 
values of~$N$, is~\hbox{displayed} in Fig.~\ref{fig:N}(c). 
The~shaded
area 
represents 
the region where the ratio of~$L$ to the  extraction
size 
$A_E^{1/2}$ is smaller than~1. This region
corresponds to
a crowding regime where
the \hbox{different}
species hinder the 
mobility of each other, thus reducing the sorting efficiency. 
This is qualitatively confirmed 
by observing the snapshots of
configurations obtained from simulations of the lattice-gas model for different
values of~$N$, where the existence of two different
regimes, a low-density one and a 
crowded  
one, can be clearly distinguished (Fig.~\ref{fig:snapN}).
A~convenient
measure of the 
mutual affinity of homotypic molecules, 
both in a dilute and 
in a crowded environment, is the effective,
macroscopic
interaction strength $C$,
which can be easily computed by inverting
Eqs.~\ref{eq:domain_formation}~and~\ref{eq:domaindensity}~\cite{ZVS+21,FPP+22}.
Fig.~\ref{fig:N}(d)
shows that $C$ 
increases monotonically as a function of the
microscopic interaction strength $g$
(as previously observed 
for
$N=1$ 
in Ref.~\onlinecite{FPP+22}),
but decreases monotonically for increasing $N$,
as predicted by the phenomenological theory.

Altogether, these numerical 
results 
suggest that 
a \hbox{typical} 
membrane system 
may in principle 
sort \hbox{simultaneously} up to $10^1$--$10^2$ 
distinct molecular species,  
with the highest value attainable only in the optimal regime, 
while
strong crowding effects
are expected to
impair
molecular sorting
for larger values of $N$.

\subsection{Sorting species with different mutual affinities}
\begin{figure} 
  \includegraphics[width=0.35\textwidth]{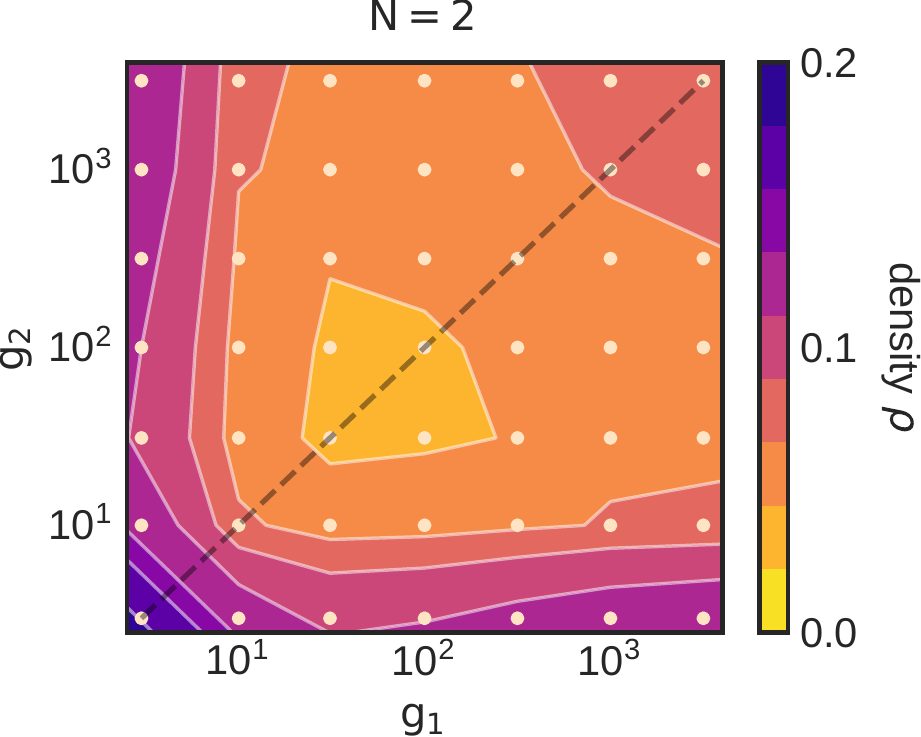}
  \caption{Optimal sorting of two distinct molecular species (lighter region) is obtained for equal homotypic affinities (dashed line $g_1=g_2$). Simulations were performed at fixed
  incoming molecule flux $\phi/k_D=10^{-5}$  
  on a square lattice. Dots correspond to 
  computed values of $\rho$, level curves are obtained by linear interpolation of the computed values.}
  \label{fig:multipleg}
 \end{figure}

For the symmetric case, where all
the molecular species have similar mutual affinities, we found simple scaling
laws for the molecular density at the steady state in the optimal region, 
away from the crowding regime.
It is then interesting to investigate to which extent this \hbox{symmetry} requirement
is restrictive.
To this aim, we~here \hbox{consider}
the process of sorting of $N=2$
molecular species, such that their mutual affinities are characterized by
independent, and possibly different microscopic interaction strengths $g_1$ and
$g_2$.
By measuring the stationary molecule density 
$\rho$, 
one may look for 
a global
minimum 
as a function of the two interaction strengths. 
Fig.~\ref{fig:multipleg} shows the existence of a single global minimum 
of the total molecule density $\rho$
for
$g_1=g_2=g_\mathrm{opt}$, suggesting that molecular sorting may be most
efficient when the distinct sorted molecular species have similar homotypic
affinities, 
and that the symmetry requirement imposed in the previous treatment 
may be not too
restrictive, as far as the optimal sorting regime is concerned.

\subsection{Sorting multivalent molecules}
\begin{figure}[b]
 \includegraphics[width=1.0\columnwidth]{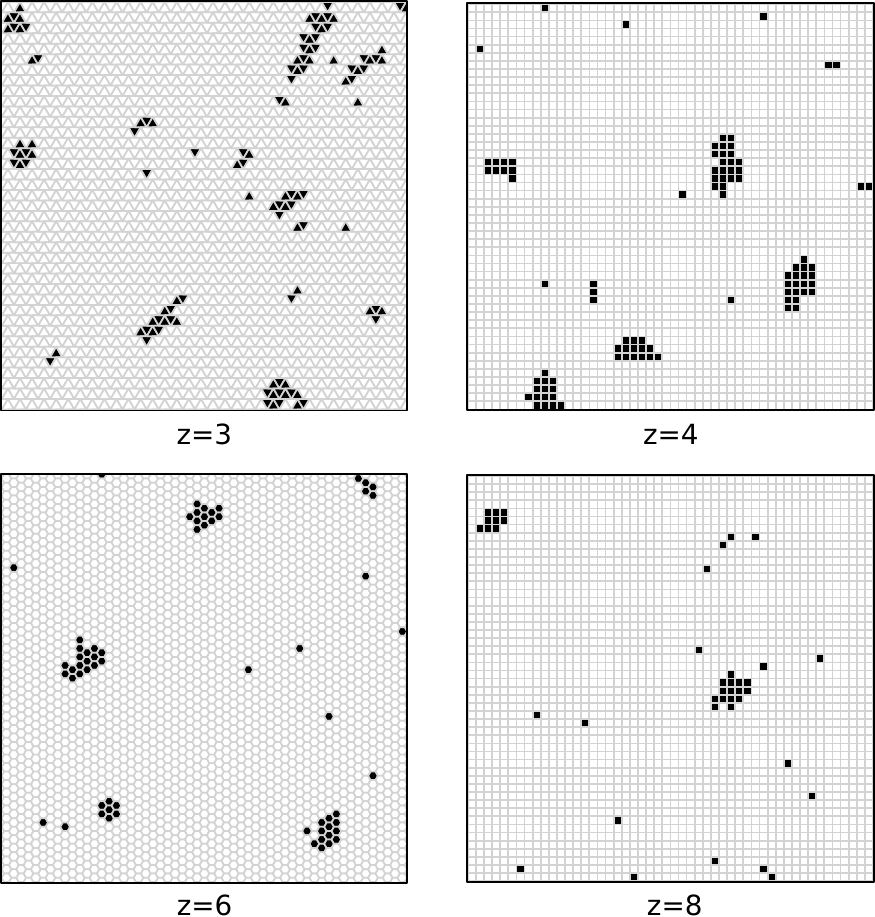}
  \caption{Snapshots of simulated optimal sorting of $N=1$ 
  molecular species 
  with incoming molecule flux $\phi/k_D=10^{-5}$ on regular lattices with different coordination number~$z$ and equal area~$A_0$ of the elementary lattice site. The panels show 
  enlargements of one quarter of the total system.}
  \label{fig:val10}
\end{figure}
\begin{figure*} 
 \includegraphics[width=0.95\textwidth]{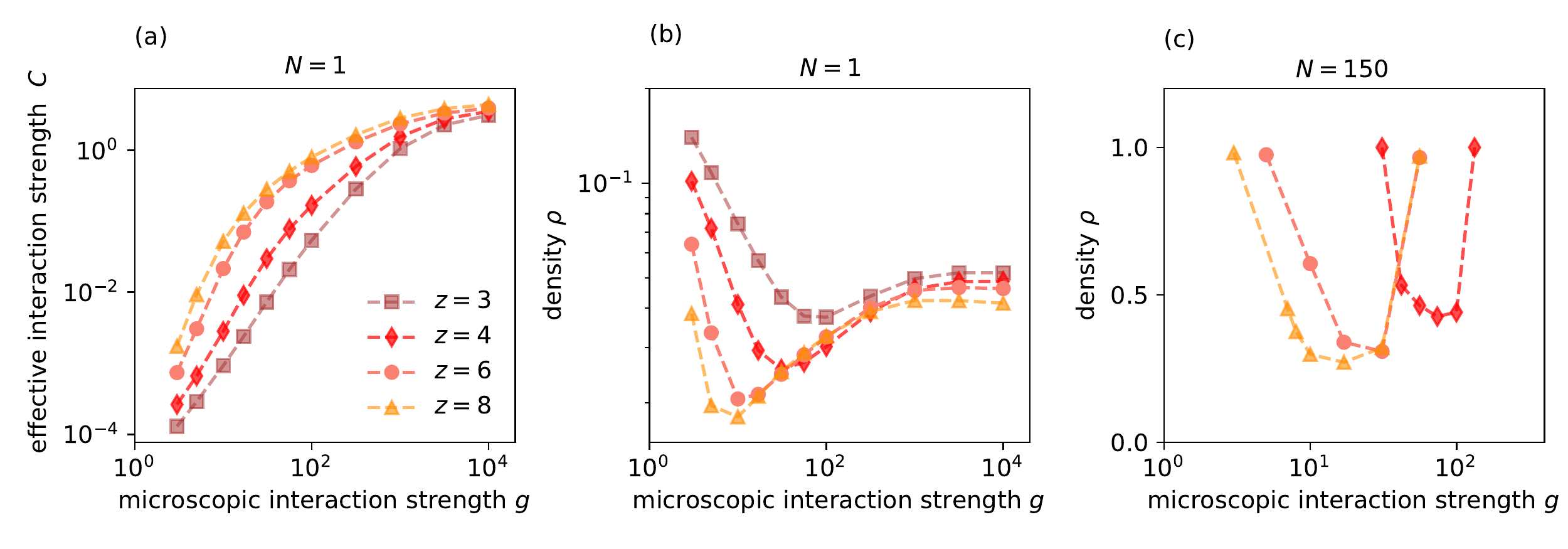}
  \caption{(a) Effective interaction strength~$C$ as a function of the
  microscopic interaction strength~$g$ and of the valence~$z$. Higher values of
  $z$ compensate for smaller~$g$. (b) Total molecule density $\rho$ as a
  function of the microscopic interaction strength~$g$ for different values of the
  valence $z$. 
  (c) Density $\rho$ as a function of the interaction strength~$g$ for different
  values of the valence~$z$ in a system with $N=150$ molecular species. Simulations were
  performed with $\phi/k_D=10^{-5}$.}
  \label{fig:val1}
\end{figure*}

An increasing amount of evidence suggests that a crucial factor in a variety of
intracellular 
phase separation 
processes is 
valence, 
which  
may be 
defined as 
the average number of 
interacting 
neighbors of a molecule
in a~phase-separated 
domain~\cite{LBC+12,BR14,SB17}.
Experiments have shown 
that 
multivalence 
promotes
domain
stability~\cite{BRP+16,LBC+12}, 
and that
multivalent protein interactions 
are responsible for
the assembly of endocytic sorting domains~\cite{DKW+21}. 

A simple way to investigate the role of valence in the 
present numerical framework is
to consider the diffusion of molecules on regular lattices with different
coordination numbers $z$, i.e., on 
triangular, square, and hexagonal lattices ($z=3,4,6$) (Fig.~\ref{fig:val10}).
The $z=8$ case can be implemented by considering the square lattice where
nearest neighbors along the diagonals are considered 
in addition to 
nearest neighbors
along the horizontal and vertical directions. 
In~this
lattice-gas framework, 
the lattice coordination number $z$ 
can 
be treated as a 
proxy 
of molecular valence.
In order to correctly compare 
the sorting \hbox{dynamics} 
on lattices  
with different coordination
numbers, the 
microscopic rates were chosen in such a way  
to provide 
the same
macroscopic diffusive dynamics 
in the continuum limit 
for all $z=3,4,6,8$ 
(see App.~\ref{app:rescaling}).
\hbox{To~focus} on the dependence of the sorting process on~$z$,
the analysis
is initially restricted
to the case where a single molecular species is sorted ($N=1$).

Fig.~\ref{fig:val1}(a) shows that
higher valence 
implies
larger values of the effective, 
macroscopic
interaction strength $C$ at fixed values of the microscopic 
interaction strength~$g$,
consistent with the intuition that molecules of higher valence can
more easily aggregate and form phase-separated domains,
and that higher valence can compensate for smaller values of the microscopic aggregation strength.
This~tendency is confirmed by
Fig.~\ref{fig:val1}(b), showing that  
optimal sorting (corresponding to the minima of the density curves)
is realized at lower values of the microscopic interaction strength $g$
for increasing $z$. 
Interestingly, the corresponding optimal values of the stationary molecule density $\rho$
also decrease for increasing $z$. 
Perhaps even more importantly, 
Fig.~\ref{fig:val1}(c) shows that 
the interval of values of $g$ such that 
the sorting process 
can 
take place for high $N$ 
(here,
$N=150$) significantly 
widens for increasing $z$:
in this condition,
sorting is impossible for $z=3$ due to molecular crowding,
but is instead possible over more then a decade of $g$ values for $z=6,8$.

Altogether, these numerical results
show that, at least in the present, highly simplified 
modeling 
framework, 
higher valence 
promotes
more efficient sorting.
It is interesting to speculate that this may also be true in the case of 
actual biological systems.

\section{Conclusion}

Biological membranes host 
a 
sophisticated 
process
of
molecule
sorting 
and 
demixing,
which is essential for the formation and maintanence of the distinct \hbox{chemical} \hbox{identies}
of diverse organelle and 
membrane regions. 
It~has been suggested~\cite{ZVS+21} that the main drivers of this
hectic distillation process are the 
phase separation of 
molecules of distinct 
species
driven by their mutual 
(\hbox{direct}~\cite{BLH+17,LBC+12,BR14} or \hbox{indirect}~\cite{GCT+05,GKL+07,SVN+12,HBF18,FPA+21,CLS+20}) interactions,
and~by the tendency of 
phase-separated molecular domains to
induce membrane bending and vesicle nucleation~\cite{BJB+18}. 
Based on these assumptions, a phenomenological theory of the process of sorting
of a single molecular cargo was developed~\cite{ZVS+21,FPP+22}.
Since molecular sorting is first of all a demixing process,
it is 
interesting to generalize the \hbox{theory} to the case where $N>1$ distinct
molecular species are sorted in parallel.
The present work is a step forward in this direction, 
in which we considered 
the simplest case
where $N$ molecular species of similar biophysical properties, 
interacting
(except for excluded volume effects) only with homotypic molecules, are sorted
simultaneously on the same membrane system.
In this case, analytical arguments and numerical simulations of a hard-core 
lattice-gas model show that, when keeping fixed the total incoming molecular
flux, the average molecule residence time and the average molecule density on 
the membrane system at the steady state increase with the heterogeneity~$N$  
of the molecular pool~as~$N^{1/2}$.
Simulations performed with biologically realistic parameter values suggest that
a large number of distinct molecular species (of the order of 10--100, depending
on the degree of optimality of the process) can be sorted in parallel on a given
membrane system without significant crowding effects.
The study of the $N=2$ case for independent values of the mutual molecular
affinities suggests that sorting may be most efficient when the distinct
molecular species 
have
similar homotypic affinity.
Lastly, motivated by the crucial 
role 
of multivalent molecules in driving 
biological phase separation~\cite{LBC+12,BR14,SB17}, we analyzed the effect of 
valence on the sorting process by simulating our sorting model on regular
lattices of varying coordination number, finding that, in this framework,
higher valence allows for more efficient sorting of a large number of distinct 
molecular species over a larger interval of mutual interaction strengths.

It would be of particular interest to check whether some of the simple, general relations
here 
described
can be actually observed
in a real cellular system.

\acknowledgments
We gratefully acknowledge useful discussions with Guido Serini, Carlo Campa,
Igor Kolokolov and Vladimir Lebedev. Numerical calculations
were made possible by a SmartData@PoliTO agreement providing access to BIGDATA
high-performing computing resources
at Politecnico di Torino,
and by
a CINECA-INFN agreement providing access to resources on MARCONI at CINECA.

\newpage
\appendix

\section{Lattice-gas model} \label{sec:latticegas}
The molecular sorting process described in the main text is 
implemented  
in terms of
a lattice-gas model, where molecules of
multiple species are distributed on a two-dimensional regular lattice with
periodic boundary conditions. 
Each lattice site can host a single molecule at most, but there is no limit to
the number of species that can populate the lattice. The current state of the
system is described by a multivariate configuration where 0 marks an empty
lattice site, while 
a number 
$i\in\{1,\ldots,N\}$
marks a molecule of the $i$-th species residing on a given site.
The state of the system
evolves according to a continuous-time Markov chain consisting of the following
three processes: 
1) molecules, whose species are chosen randomly from a set of $N$ species in
such a way that the overall flux 
has the assigned value $ \phi$, 
are inserted
into empty sites with rate $k_I$; 2) molecules jump towards empty
neighboring sites with rate $k_D n_e/g^{n_h}$, where $n_e$ is the number
of empty neighboring sites, $g$ is the intermolecular interaction strength, and
$n_h$~is the number of 
homotypic
molecules in 
neighboring sites; 3) 
the molecules in a 
connected
domain of 
homotypic
molecules 
are 
extracted 
when 
the domain 
reaches the
extraction size $m$. 
(The extraction mechanism 
adopted here 
is a slightly simplified version of 
the one used in Ref.~{\onlinecite{ZVS+21}},
where 
the molecules of a connected domain were extracted only when
the domain 
grew to the point of containing
a 
molecule-filled square of
given size). A~schematic representation of these mechanisms is shown in
Fig.~\ref{fig:scheme}. 
In the simulations,
$A_0=1$, 
i.e.,
\hbox{areas} are measured as multiples of the elementary lattice
area~$A_0$,  
and the realistic value $m=25$ is used. 
This way, the relevant
microscopic parameters describing the process are the intermolecular interaction
strength $g$, the valence $z$, and the ratio $k_I/k_D$. 
For low values of the molecule
density~$\rho$, such as those experimentally measured in Ref.~\onlinecite{ZVS+21}, the 
molecule flux $\phi = k_I (1-\rho)$ 
is
approximately equal to the
insertion rate $k_I$. Compatibly with experiments al observations~\cite{ZVS+21},
simulations were performed
with $\phi/k_D =10^{-5}$.
 
\section{Crowding effects}
\label{app:crowding}
\begin{figure}
     ~\!\!\!\!\!\!\!\!\!\includegraphics[width=1.07\columnwidth]{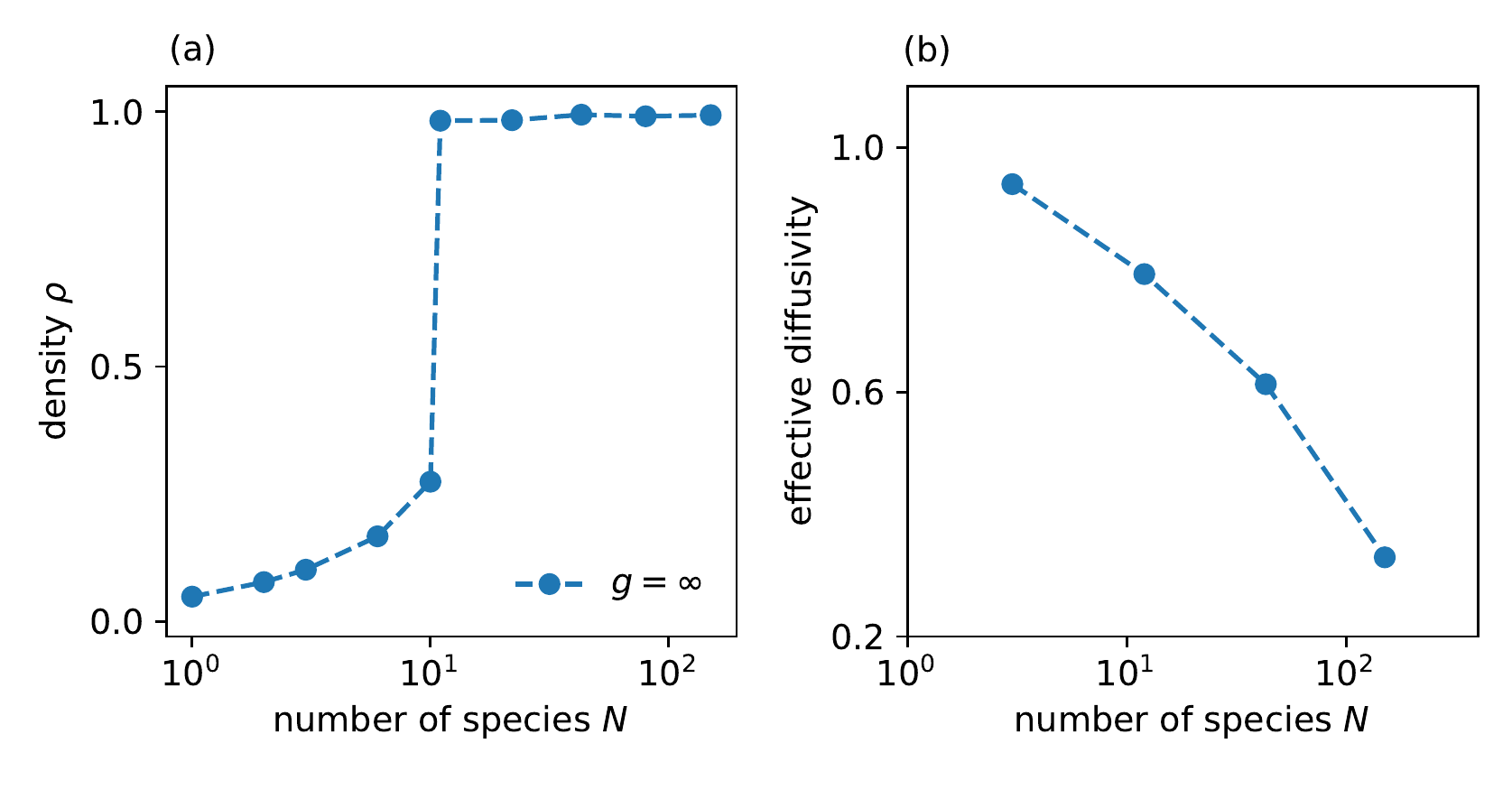}
    \caption{(a) Total molecule density $\rho$ as a function of the number of
    species $N$ 
     in the limit case
   $g=\infty$.
   Above a
    \hbox{critical} value $\sim 10$ of the number of species, the density 
    \hbox{becomes}~$\sim 1$,
    signaling that
    most
    lattice sites are filled
    and that the 
    distillation process has come to a stop due to molecule \hbox{overcrowding}.
    (b)~\hbox{The~effective} diffusivity 
    of a test particle
    decreases as the number of species $N$ increases, signaling that 
    molecular mobility is strongly reduced. Simulations were performed at fixed
  incoming molecule flux $\phi/k_D=10^{-5}$  
  on a square lattice.} 
    \label{fig:g_inf}
\end{figure}

The deviation from the $N^{1/2}$ scaling observed at large $N$ 
in Fig.~\ref{fig:N}(b)
is a consequence
of the breakdown 
of the low-density regime where 
the processes of sorting of the different molecular species
are approximately independent~(Fig.~\ref{fig:g_inf}(a)).
For large $N$, 
the sorting
domains are no longer well separated and 
molecular 
mobility 
is strongly
reduced. 
This~can be checked 
by 
tracking 
the diffusive motion of a test molecule 
which does not interact with any of the molecules of 
the 
$i=1,\ldots,N$ 
species,
except for excluded volume effects. 
The effective diffusivity of the test molecule, 
measured empirically
from  
the temporal growth of its mean squared displacement,
is observed to decrease when the number $N$ of molecular species increases~(Fig.~\ref{fig:g_inf}(b)).

\section{Relation between microscopic and macroscopic diffusivities}
\label{app:rescaling}

\begin{figure}[b!]
     \includegraphics[width=\columnwidth]{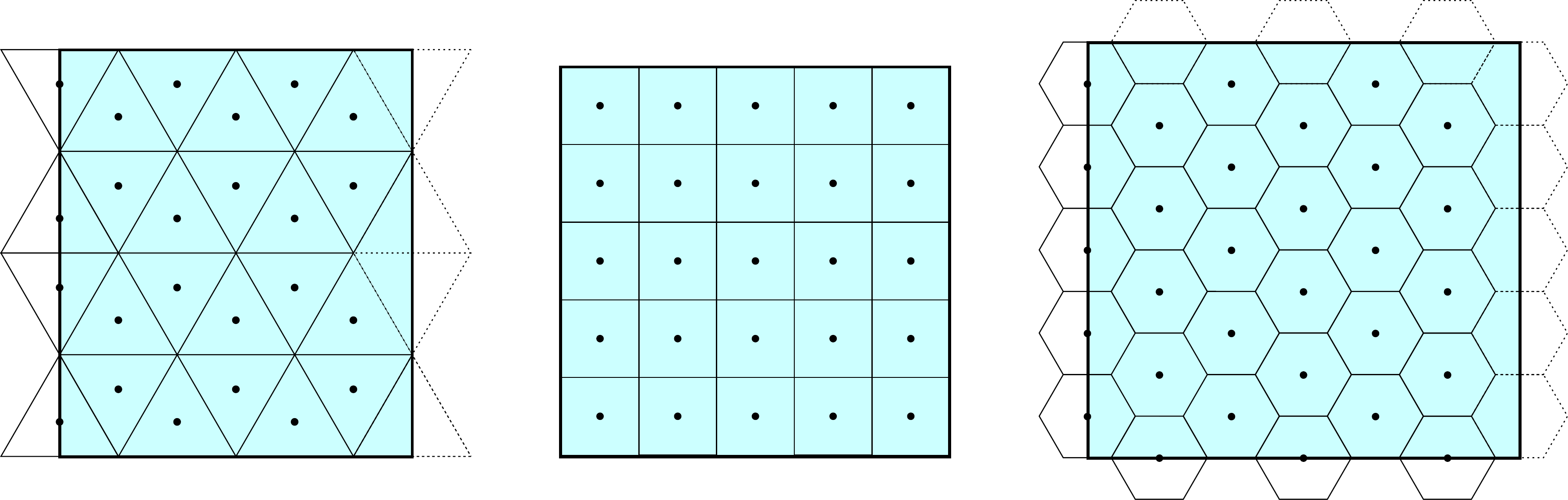}
    \caption{Small periodic regular lattices with $z=3,4,6$, with identical
      area $A_0$ of the elementary lattice site and approximately equal total area (light
      blue). The centers of the sites belonging to each lattice are
      marked with a black dot. } 
    \label{fig:reticoli}
\end{figure}

The microscopic rate $k_D$ of jump to an empty neighboring site of the lattice
can be related to the macroscopic diffusivity $D$ 
as follows.
Consider the diffusion of a single molecule 
on an otherwise empty lattice,
and let $n_x (t)$ be the probability that the molecule occupies site $x$ at
time~$t$. At time $t + \delta t$,
\begin{equation}
  n_x (t + \delta t) = n_x (t) + k_D\, \delta t \cdot \sum_{y \in \partial x}
  (n_y - n_x) \label{eq:dprob}
\end{equation}
where $\partial x$ is the set of nearest neighbors of $x$. 
For
the
regular lattices with $z = 3, 4, 6$, expanding $n_y$ in a Taylor series
centered in $x$ and dividing by $\delta t$, Eq.~\ref{eq:dprob} tends to the
diffusion equation for $n_x$, with 
$D = \frac{1}{4}\,z\, k_D\, d^2$, 
where~$z$ is the
number of nearest neighbors of $x$, and $d$ is the distance between the center
of neighboring sites. 
The same procedure applied to 
the $z = 8$ case with diagonal neighbors 
gives instead
$D = 3\, k_D\, d^2$.
\newpage
For a correct comparison of the sorting dynamics 
in the presence of 
different coordination
numbers, 
regular lattices of different coordination number
but identical area~$A_0$ of the elementary lattice site 
(i.e., of identical area per molecule) were used
(Fig.~\ref{fig:reticoli}). The total area of each lattice was chosen 
to contain
approximately $100^2$ sites, and 
the microscopic jump rate $k_D$ was rescaled to provide the same
value of the 
macroscopic diffusivity $D$ for all $z=3,4,6,8$.

\end{document}